# Natively Fat-Suppressed 5D Whole-Heart MRI with a Radial Free-Running Fast-Interrupted Steady-State (FISS) Sequence at 1.5T and 3T

Jessica AM Bastiaansen[1], Davide Piccini[1,2], Lorenzo Di Sopra[1], Christopher W Roy[1], Robert R Edelman[4,5], Ioannis Koktzoglou[4,6], Jerome Yerly[1,3], Matthias Stuber[1,3]

[1]Department of Diagnostic and Interventional Radiology, Lausanne University Hospital and University of Lausanne, Lausanne, Switzerland

[2] Advanced clinical imaging technology, Siemens Healthcare AG, Lausanne, Switzerland

[3]Center for Biomedical Imaging, Lausanne, Switzerland

[4]Department of Radiology, NorthShore University HealthSystem, Evanston, Illinois, USA.

[5]Northwestern University Feinberg School of Medicine, Chicago, Illinois, USA.

[6]The University of Chicago Pritzker School of Medicine, Chicago, Illinois, USA.

**To whom correspondence should be addressed:** Jessica AM Bastiaansen, Department of Radiology, University Hospital Lausanne (CHUV), Rue de Bugnon 46, BH 8.84, 1011 Lausanne, Switzerland, Phone: +41-21-3147516, Email: jbastiaansen.mri@gmail.com

**Category:** Rapid Communication

**Word count manuscript (body text):** 3494

**Figure count:** 7

**Table count:** 0

**Supporting figure count:** 5

**ABSTRACT**

**Purpose:** To implement, optimize and test fast interrupted steady-state (FISS) for natively fat-suppressed free-running 5D whole-heart MRI at 1.5T and 3T.

**Methods:** FISS was implemented for fully self-gated free-running cardiac- and respiratory-motion-resolved radial imaging of the heart at 1.5T and 3T. Numerical simulations and phantom scans were performed to compare fat suppression characteristics and to determine parameter ranges (readouts per FISS module (NR) and repetition time (TR)) for effective fat suppression. Subsequently, free-running FISS data were collected in ten healthy volunteers. All acquisitions were compared with a continuous bSSFP version of the same sequence, and both fat suppression and scan times were analyzed.

**Results**: Simulations demonstrate a variable width and location of suppression bands in FISS that was dependent on TR and NR. For a fat suppression bandwidth of 100Hz and NR≤8, simulations demonstrated that a TR between 2.2ms and 3.0ms is required at 1.5T while a range of 3.0ms to 3.5ms applies at 3T. Fat signal increases with NR. These findings were corroborated in phantom experiments. In volunteers, fat SNR was significantly decreased using FISS compared with bSSPF ($p<0.05$) at both field strengths. After protocol optimization, high-resolution (1.1mm$^3$) 5D whole-heart free-running FISS can be performed with effective fat suppression in under 8 min at 1.5T and 3T at a modest scan time increase compared to bSSFP.

**Conclusion**: An optimal FISS parameter range was determined enabling natively fat-suppressed 5D whole-heart free-running MRI with a single continuous scan at 1.5T and 3T, demonstrating potential for cardiac imaging and noncontrast angiography.

**Keywords** (3 to 10): Fat suppression, FISS, bSSFP, radial, whole-heart, 1.5T MRI, 3T MRI, noncontrast.

**List of abbreviations**: FISS, fast interrupted steady state; bSSFP, balanced steady state free precession; RF, radio frequency; MRI, magnetic resonance imaging; FS, fat suppression; NR, Number of readouts per FISS module; TR, repetition time; MRA, magnetic resonance angiography;

## INTRODUCTION

Balanced steady-state free precession (bSSFP) imaging techniques are widely used for cardiovascular imaging since they provide a high signal-to-noise ratio (SNR), and a high blood-myocardium contrast. (1–3). A drawback of bSSFP is its sensitivity to flow artifacts and $B_0$ inhomogeneities (4), which are more pronounced at higher magnetic field strengths. Moreover, bSSFP sequences require fat suppression, respiratory and cardiac motion compensation, long scan times, specific slice prescriptions, and complex patient setups to adequately visualize small blood vessels, especially the coronary arteries (5).

Whole-heart imaging approaches that are free-breathing and non-ECG-triggered (i.e. free-running) are gaining in popularity because they effectively reduce scanning complexity (6,7) by removing the need for specific slice orientations, gating, triggering, and ECG lead placement. Cardiac images can then be retrospectively reconstructed in arbitrary spatial and temporal dimensions and provide information on both cardiac anatomy and function from one single acquisition. The ideal free-running approach should: (a) provide high signal and contrast, (b) perform at different magnetic field strengths, (c) provide high spatial and temporal resolutions, (d) provide adequate fat signal suppression, and (e) be time efficient.

In current free-running approaches, the manner in which high SNR and contrast are generated is field strength dependent with bSSFP being used for applications at 1.5T (6), and gradient-echo (GRE) sequences in combination with contrast agents at 3T (7). Both methods achieved high spatial and temporal resolution using flexible radial sampling. As for fat suppression, at 1.5T this was performed using fat saturation pulses that repeatedly interrupted the steady-state magnetization in bSSFP (6). However, fat saturation in combination with mandatory ramp-up pulses remains relatively time inefficient and the water-fat cancellation artifact could still be observed at some of the coronary vessel borders (6). The free-running approach at 3T used water excitation pulses to suppress fat signals (7). Shorter scan durations may be obtained by the use of advanced reconstruction methods such as XD-GRASP (8). Its application to data acquired with free-running bSSFP (6), referred to as whole-heart motion-resolved 5D MRI (9), significantly improved cardiac image quality over the use of motion-correction methods at 1.5T but the scan durations remained over 14min (6,9). Post contrast, at 3T, and without the use of compressed sensing reconstruction, scan durations of 10min have been reported (7). Therefore, there is currently a need for an MRI sequence that provides high tissue contrast, effective fat signal suppression, and improved time efficiency within the free-running framework, while the use of contrast agent should remain an option, but not a requirement.

The relatively long scan durations in previous free-running bSSFP studies are mainly caused by the mandatory ramp-up pulses applied after each fat saturation pulse (6,9), and in free-running GRE by lengthy water excitation pulses resulting in longer TRs (7). Compared to fat suppression pulses (10), water excitation pulses (11) would not interrupt the steady-state in bSSFP but may increase the repetition time (TR). Even the use of relatively short water excitation pulses such as LIBRE (12–14) would prolong the TR by some extent.

Recently, a 2D radial fast interrupted steady-state (FISS) sequence has been developed that provides the high SNR and blood-muscle contrast-to-noise (CNR) ratio of bSSFP while simultaneously suppressing fat under certain conditions (15). FISS demonstrated a reduction in both flow artifacts compared with bSSFP, and decreased arterial signal saturation relative to GRE (15,16). Another advantage of FISS with respect to bSSFP include reduced signal oscillations from off-resonant tissues when approaching the steady-state (17). In a free-running context, a FISS implementation may lead to intrinsically fat-suppressed imaging without needing periodically applied fat suppression and ramp-up pulses, which will be conducive of improved time efficiency and visualization of anatomical structures embedded in epicardial fat.

While 2D radial implementations of FISS have originally been reported, the above advances the hypothesis that a carefully optimized 3D version of FISS may be exploited for fat-suppressed respiratory and cardiac motion-resolved free-running whole-heart imaging. Therefore, the purpose of this study was to 1) develop and implement a 3D radial FISS sequence as part of a respiratory and cardiac self-gated free-breathing cardiac and respiratory motion-resolved 5D imaging framework, to 2) provide a thorough analysis of native fat suppression capabilities of FISS within this framework by simulations and phantom experiments, 3) provide parameter ranges for free-running, fat-suppressed imaging to be effective at 1.5T and 3T, and to 4) demonstrate its feasibility for natively fat-suppressed whole-heart imaging at both field strengths in a single continuous scan and while comparing results with bSSFP.

**METHODS**

*Theory*

FISS sequences frequently interrupt the bSSFP echo train (15) and use these fast interruptions to induce broad signal suppression bands, the location of which - and bandwidth of which - are highly dependent on the TR and the frequency of the interruptions. The fat-suppressing properties of FISS and the corresponding sequence parameters were thoroughly investigated in this study.

A typical FISS module starts with a α/2 tip-down pulse, is followed by RF excitation pulses of which the number depends on the amount of readouts per FISS module (NR), and ends with a α/2 tip-up pulse (Figure 1a) (15). Each FISS module is followed by gradient spoilers, RF pulses are phase alternated by 180°, and RF spoiling is applied across different FISS modules using a 117° phase increment (15). The repeated interruptions cause a sequence duration increase compared with bSSFP, especially for lower values of NR, and can be estimated as follows:

$$t_{sequence} \approx (N_{readouts}/NR) * ((NR + 1) * TR + t_{pulse} + t_{spoiler})$$

With $N_{readouts}$ defining the total number of readouts in a sequence, $t_{spoiler}$ the duration of the spoiler gradient, and $t_{pulse}$ the RF excitation pulse duration. Sequence durations of bSSFP sequences can be estimated using an NR that approaches $N_{readouts}$.

*Pulse sequence implementation for 3D radial free-running FISS*

A prototype 3D radial FISS sequence was implemented and integrated into a 5D whole-heart sparse MR imaging methodology (9) subsequently developed into a fully self-gated free-running framework (18). FISS was programmed as a continuous sequence without magnetization preparation modules (Figure 1b).

As a basis for the FISS implementation, a free-running bSSFP sequence was used, which is a 3D radial sequence (6) that follows a spiral phyllotaxis k-space sampling pattern (19) that is rotated about the z-axis by the golden angle (20)(Figure 1c). The first readout of each radial interleave is oriented along the superior-inferior (SI) direction and is used for cardiac and respiratory self-gating.

The amount of segments (i.e., k-space readouts per radial interleave) was fixed to keep a similar radial trajectory as in (6,9). Each 3D radial interleave contained a total of 24 segments that can be divided into a flexible number of FISS modules that allowed for NR=1,2,3,4,6,8,12,24. Non-slice-selective rectangular RF excitation pulses of 0.3ms duration were used as opposed to earlier reported thin slab selective RF excitation pulses (15). A scan duration comparison was made between FISS and bSSFP using the vendor provided sequence simulator and a TR of 3.0ms.

*Numerical simulations*

Numerical simulations of the Bloch equations were performed in Matlab (The MathWorks, Inc., Natick, MA, United States) to compute the transverse magnetization ($M_{xy}$) of FISS and bSSFP sequences for a range of resonance frequencies and sequence parameters, providing a quantitative comparison of signal suppression and suppression bandwidth. Unless otherwise stated, simulations included rectangular RF excitation pulses of 0.3ms duration, RF excitation angles of 40°, perfect RF and gradient spoiling between subsequent FISS modules by nulling $M_{xy}$, a $T_1$ of 1600ms and a $T_2$ of 180ms as used before in (15).

A first set of simulations was performed to compare $M_{xy}$ (i.e. measured signal) of FISS and bSSFP sequences as a function of tissue frequency and RF excitation angle. The tissue frequencies varied between -500Hz and 100Hz (steps of 2Hz), the RF excitation angles between 20° and 60° (steps of 1°), the NR was set to 1, and the TR to 3.0ms.

In a second set of simulations, $M_{xy}$ was determined as a function of TR and tissue frequency to investigate the suppression bandwidth, defined as the bottom 10% (i.e. a 10-fold reduction of fat signal) of the maximum on-resonant water $M_{xy}$. Tissue frequencies were varied between -500Hz and 100Hz (steps of 1Hz) and the TR between 2.0ms and 4.0ms (steps of 0.05ms). Simulations were repeated for NR=1,2,3,4,6,8,12,24.

As these two simulations illustrate the FISS signal response only for precise tissue frequencies a third set of simulations was performed to quantify a more realistic fat signal suppression as a function of TR and NR at 1.5T and 3T. Here, the fat signal was defined as a normalized Gaussian spectrum with a full-width-at-half-maximum (FWHM) of 50Hz centered around -220Hz (1.5T) or -440Hz (3T). The $M_{xy}$ of this spectrum was simulated by varying the TR from 2.0ms to 4.0ms (in steps of 0.05ms), and by varying the NR from 1 to 24 (in steps of 1).

### *Phantom experiments*

Fat suppression capabilities were tested in baby oil (Johnson and Johnson, New Brunswick, NJ) on a clinical 1.5T and 3T MRI system (MAGNETOM Aera and MAGNETOM Prisma[FIT], Siemens Healthcare, Erlangen, Germany), with a field-of-view of (220mm)$^3$, matrix size of 112$^3$, RF excitation angle of 40°. The TR was varied from the minimum TR possible, from 2.3ms (1.5T) and 2.5ms (3T), to 4.0ms in steps of 0.1ms by varying the receiver bandwidth. Experiments were repeated for NR=1,2,3,4,6,8,12,24. In a second experiment, with a FISS scan (NR=8, TR=3.0ms) the effect of gradients spoilers was tested as they contribute significantly to the scan duration

### *5D Whole-heart free-running FISS and bSSFP in volunteers*

A free-breathing non-ECG-triggered 3D radial FISS sequence was used to perform whole-heart imaging in ten volunteers. All experiments followed institutional guidelines and all volunteers provided written informed consent. In each volunteer a FISS and a bSSFP acquisition were performed, either at 1.5T or 3T (5 volunteers on each scanner). Imaging parameters for both sequences were as follows: field-of-view of (220mm)$^3$, RF excitation angles of 40-45° (3T) and 50-60° (1.5T), depending on SAR limitations, but always the same for both scans. Whole-heart volumes were acquired with an isotropic spatial resolution of 1.1mm$^3$ or 1.38mm$^3$, with respective matrix sizes of 192$^3$ and 160$^3$, with the acquisition of 3419 or 5181 radial interleaves (with each 24 segments), resulting in 82,056 or 124,344 acquired k-space lines. For experiments at 1.5T, the lowest possible TR was used which varied from 2.75ms to 3.0ms and depended on the spatial resolution. At 3T, the TR was varied between 3.0ms and 3.4ms.

*Motion extraction and motion-resolved image reconstruction*

The acquired 3D data were sorted into a 5D motion-resolved dataset that contains separate respiratory and cardiac dimensions as described previously (8,9). Both respiratory and cardiac motion signals were extracted from the SI projections (19) acquired at the beginning of each radial interleave (Fig. 1b)(18). The self-gating signals were used for data binning in the respiratory dimension, with 4 motion states resolved from end-inspiration to end-expiration, and in the cardiac dimension, with cycles temporally divided into windows of 50ms width. 5D motion-resolved images were then reconstructed using XD-GRASP (8).

*Data analysis*

Phantom data and non-motion-resolved volunteer data, reconstructed at the scanner, were used to quantify SNR in compartments containing oil or chest fat by dividing the average signal of fat by the standard deviation of the background signal outside the region of interest.

A cardiac volume at end-expiration and mid-diastole was selected for visual comparison of coronary vessel reformats obtained with SoapBubble(21).

A paired Student's t-test was performed on phantom and volunteer data. Results are represented as average ± one standard deviation and $p<0.05$ was considered statistically significant.

**RESULTS**

*FISS and bSSFP scan duration*

The acquisition time of FISS is highly variable and depends on NR. Using a single readout per FISS module (NR=1), the acquisition time of a single radial interleave increases by 205% compared to a non-fat-suppressed bSSFP sequence (Figure 1d). By increasing NR to 24, the relative acquisition time increase is 8% compared with bSSFP. Although high NR values may be favorable for scan time duration, it affects fat suppression behavior. The relative scan time increase comparing FISS and non-fat-suppressed bSSFP for NE=4 and NE=8 is 53% and 26% respectively (Figure 1d).

*Numerical simulations*

FISS provides the same maximum on-resonant (~0Hz) signal intensity compared with bSSFP for a similar range of RF excitation angles, but with FISS a broad suppression bandwidth can be observed around -220 Hz and -440 Hz (Supporting Figure S1).

Numerical simulations show the effect of TR and tissue frequency on the signal behavior for NR=1, 4 and 8 (Figure 2, see also Supporting Figure S2 for the full range of NR that was simulated). NR influences the number of relatively narrow bands (with FWHM of ~10-20Hz) that disrupt the suppression bandwidth (Figure 2a,d,g).

In practical terms, concomitantly with increasing NR, these bands increase in frequency, which decreases the signal suppression bandwidth while increasing the TR changes the location of the suppression bands. Increasing NR will reduce scan time on the one hand but also decreases the range of TRs that are adequate for fat suppression.

Since the ideal sequence would provide broad fat suppression with reasonable scan times, and considering the above findings, the following parameter combination is proposed for use in 5D whole-heart free-running imaging: With NR=8, at 1.5T, a TR in the range of 2.0ms to 3.0ms (~80Hz suppression bandwidth)(Figure 2g,h) should be chosen. At 3T, a TR in the range of 3.0ms to 3.5ms (~60Hz bandwidth) (Figure 2g,i) was deemed optimal.

Using a Gaussian fat signal distribution mitigates the effect of the ~10-20Hz excitation bands occurring at specific frequencies, and contributes to a more intuitive understanding of applicable TR values. When using a Gaussian fat signal distribution, it can be seen that the total contributing fat signal at 3T drops to below 10% for TRs around 3.2ms (Figure 3a). It also demonstrates that with NR=4, fat suppression may be achievable with TRs of 2.7ms, compared with 3ms using NR=2 or 3 (Figure 3b). Revealing that a FISS acquisition of NR=4 poses a better candidate compared to NR=2 or 3, allowing even shorter scan durations. At 1.5T, the range of possible TRs is wider when compared to that at 3T (Figure 3d), and also here, the simulation suggests that a FISS acquisition with NR=4 performs better compared with NR=3 or 2 (Figure 3e).

*Phantom experiments*

Phantom scans demonstrated a clear difference in fat suppression of FISS when compared to bSSFP, with FISS approaching bSSFP behavior with increasing NR. Fat SNR behavior in phantom experiments corresponded well with that of the simulations at both field strengths (Figure 3be & cf).

No change in image quality was observed with and without the use of gradient spoilers (Supporting Figure S3a,b), but a clear signal fluctuation in SI projections was observed when spoilers were not applied (Supporting Figure S3c,d). SI projections form a necessary component for accurate motion extraction and thus gradient spoilers must be included.

*5D Whole-heart free-running FISS and bSSFP in volunteers*

Free-running FISS acquisitions were successfully performed in all volunteers. In general, a large difference in image quality was observed when comparing FISS to bSSFP (Figure 4-7), with the former demonstrating markedly reduced streak artifact. In all volunteers overall fat SNR on FISS images relative to bSSFP, was significantly decreased ($p<0.01$) from $62.4 \pm 30.0$ to $25.6 \pm 17.0$ (Figure 7b).

The use of the free-running framework enabled a clear visualization of the different respiratory and cardiac motion states using both FISS and bSSFP acquisitions (Figure 4). On FISS images fat signal originating from the chest was visibly lower when compared with bSSFP (Figure 4-7) which is consistent with the reduced level of streaking artifacts.

Coronary arteries were also successfully visualized in some of the reconstructed cardiac phases in coronal views (Figure 5-6) and can be better delineated in the reformats (Figure 7a) where a strong water-fat cancellation artifact can be observed at the vessel borders using bSSFP but not using FISS, and independent of the magnetic field strength. The scan durations of the free-running FISS acquisitions depended on NR, TR and resolution, and varied between 4:46min (NR=8, TR=2.75, 1.38mm$^3$ resolution, 3:45min for bSSFP) and 9:26min (NR=4, TR=3.0ms, 1.1mm$^3$ resolution, 6:19 for bSSFP). Note, that using the free-running framework, scan time is no longer dependent on respiration or cardiac frequency.

## DISCUSSION

In this study a 3D radial FISS sequence was implemented as part of a 5D whole-heart free-running framework. The capabilities of the sequence in terms of effective fat suppression with respect to bSSFP was demonstrated in simulations and phantoms, and a full range of sequence parameters for fat suppression at both 1.5T and 3T was provided. Consistent with these findings, 5D whole-heart free-running FISS showed promising results for motion-resolved whole-heart imaging and its potential for nonenhanced MRA at both 1.5T and 3T that comes with only a modest increase in scan time compared with non-fat-suppressed bSSFP. For comparison, and if fat saturation and ramp-up pulses were inserted to provide for fat suppression in bSSFP, the scan time would be 14min (6,9), which is 80% higher than that of FISS (7min49) with NR=8 and a TR of 3ms.

The FISS protocol was optimized in simulations using NR values up to 24, and the relationship between this parameter and fat suppression bandwidth was elucidated in detail. Phantom results confirmed that fat signal suppression is effective for a wide range of NR, and showed decreased fat suppression compared to bSSFP. Clearly, the range in TR for which effective fat suppression can be obtained is constrained. However, using the plots derived from numerical simulations and phantom experiments in this study, it is now possible to prescribe the TR that leads to maximized in vivo fat suppression for different NRs and for both 1.5T and 3T.

The free-running FISS protocol was optimized based on simulations and achieved a reduced fat signal and reduced streak artifact following a motion-resolved reconstruction approach when compared to an equivalent free-running bSSFP protocol without fat suppression. Clearly, the level of fat suppression obtained with FISS enables the visualization of small anatomical structures such as the coronary arteries. FISS also produced comparable cardiac images with respect to bSSFP in terms of signal and motion, without using contrast agents. In the current study, the occurrence of water-fat cancellation artifacts with bSSFP could be clearly observed at the vessel borders. When fat saturation is inadequate, water-fat cancellation artifacts may still be observed, as was the case in some of the vessel borders displayed in previous fat-suppressed free-running bSSFP approaches at 1.5T (6,9) but not on FISS as investigated here. In terms of scan time compared with previous free-running approached, the GRE free-running approach at 3T had a reported scan duration of 10 min and necessitated the use of contrast agents, while that of the fat-suppressed bSSFP counterpart at 1.5T led to 14min scanning time. Therefore, the current free-running FISS implementation may provide an effective alternative at both field strengths.

In the recently reported 2D radial cine FISS sequence protocols, NR values ranging from 1 to 8 were used (15,16). In the current study, 5D whole-heart free-running FISS was performed using either NR=4 or NR=8. In our specific use case, this divides the spiral interleave consisting of 24 lines into 6 or 3 FISS modules. A limitation of FISS is that it still increases scan time compared with a continuous albeit non-fat-suppressed bSSFP acquisition. Future studies may investigate the utility of higher NR values, or a lower number of acquired k-space lines per spiral interleave, which were not investigated in this study but may further decrease scan time. The application of a gradient spoiler increased the scan time but was critical for the acquisition of stable SI projections aimed at a total self-gated respiratory and cardiac motion extraction. The gradient spoiler strength and duration was not optimized but this may help reduce the scan time further. Using pilot-tone navigation (22) motion extraction would no longer rely on stable SI projections and may be omitted from the sequence together with the gradient-spoilers. Lastly, further improvements in fat suppression may be achieved by tuning the TR in a patient-specific way, by incorporating a detailed fat spectrum analysis.

The proposed 5D whole-heart free-running FISS approach offers native fat suppression and reduced streak artifact with a modest (~25%) increase in scan duration when compared to non-fat-suppressed bSSFP. Cardiac MRI examinations are typically challenging because of the complex anatomy of the heart, the epicardial fat, and both the need for large volumetric coverage and adequate motion suppression. Using a 5D motion-resolved imaging approach that incorporates fat suppression, these issues can be addressed, and patient set-up time and scan planning are substantially abbreviated as both ECG lead placement and scan plane positioning are no longer needed.

## CONCLUSION

3D radial FISS was implemented and optimized using numerical simulations and phantom studies, and demonstrated highly effective fat suppression and reduced streak artifact with modest scan time increases compared to non-fat-suppressed bSSFP. The integration of 3D radial FISS in a free-running framework enabled natively fat-suppressed fully self-gated respiratory and cardiac motion resolved whole-heart imaging at both 1.5T and 3T, and its feasibility for non-contrast coronary angiography was demonstrated.

## ACKNOWLEDGEMENTS

JB received funding from the Swiss National Science Foundation (grant number PZ00P3_167871), the Emma Muschamp foundation, and the Swiss Heart foundation. MS received funding from the Swiss National Science Foundation (grant number 173129, 150828, and 143923). RRE received funding from the National Institutes of Health (grant number R01 HL130093 and R01 HL137920). IK received funding from the National Institutes of Health (grant number R01 EB027475).

Author contributions were as follows: JB designed the study, implemented the sequence, performed the simulations, acquired and analyzed the data and wrote the manuscript. DP contributed the sequence code for free-running and contributed to the study design. LDS provided the self-gating code, CR contributed to the data acquisition, IK and RRE provided sequence code snippets for FISS, JY provided the 5D reconstruction framework, MS contributed to the study design and drafting of the manuscript. All authors read and revised the manuscript.

number R01 HL130093 and R01 HL137920). IK received funding from the National Institutes of Health (grant number R01 EB027475).

Author contributions were as follows: JB designed the study, implemented the sequence, performed the simulations, acquired and analyzed the data and wrote the manuscript. DP contributed the sequence code for free-running and contributed to the study design. LDS provided the self-gating code, CR contributed to the data acquisition, IK and RRE provided sequence code snippets for FISS, JY provided the 5D reconstruction framework, MS contributed to the study design and drafting of the manuscript. All authors read and revised the manuscript.

**FIGURE CAPTIONS**

*Figure 1*

Sequence diagram illustrating the components within a FISS module (a). Each FISS module starts with a tip-down pulse of α/2, is followed by RF excitation pulses of which the number depends on the amount of readouts per FISS module (NR) and ends with a tip-up pulse α/2. After each FISS module the magnetization is spoiled and each RF pulse, regardless of NR, is phase alternated by 180°. RF spoiling by 117 degrees is applied across FIS modules. Within a free-running framework, FISS is implemented as a free-breathing and non-ECG-triggered sequence without application of magnetization preparation modules (b). In the current illustration, each 3D radial spiral interleave is composed of 6 FISS modules (NR=4), and each spiral is rotated by the golden angle (c). The stop and restart of the FISS sequence will lengthen the acquisition of a single 3D radial interleave compared to the use of a continuous uninterrupted bSSFP sequence (d).

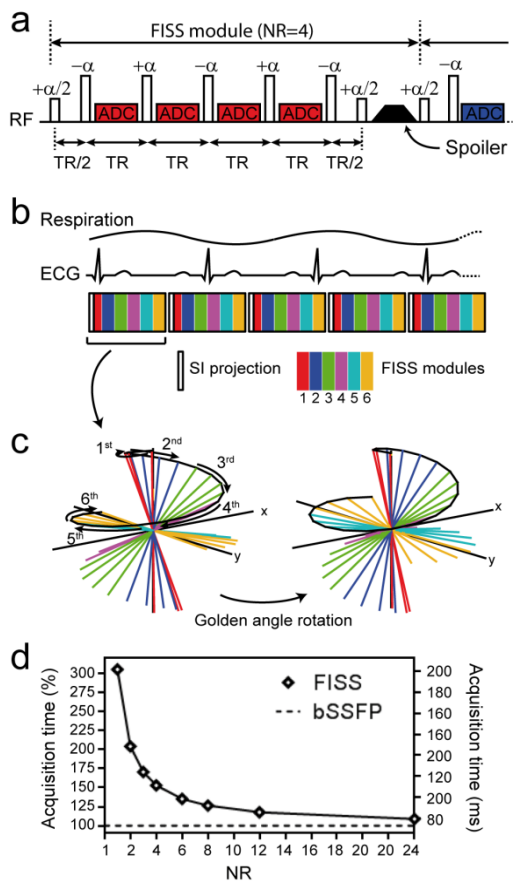

*Figure 2*

Bloch simulations of FISS illustrate the effect on the transverse magnetization as function of TR and tissue frequency and simultaneously depicts the behavior for fat at 1.5T (-220Hz) and 3T (-440Hz). For NR=1 (a,b,c), NR=4 (d,e,f) and NR=8 (g,h,i) different signal behavior can be observed. An increase in the number of readouts per FISS module (NR) mainly affects the width of the fat suppression bands, while a change in TR mainly affects the location of the bands. Simulations up to NR=24 were performed (See Supporting Figure 2). The fat resonances at 1.5T and 3T are indicated by the vertical lines in (a,d,g). The horizontal green lines (in a,d,g) correspond to fat signal responses at 1.5T that match TR values (in b,e,h) and the red lines correspond to fat signal esponses at 3T that match TR values (in c,f,i).

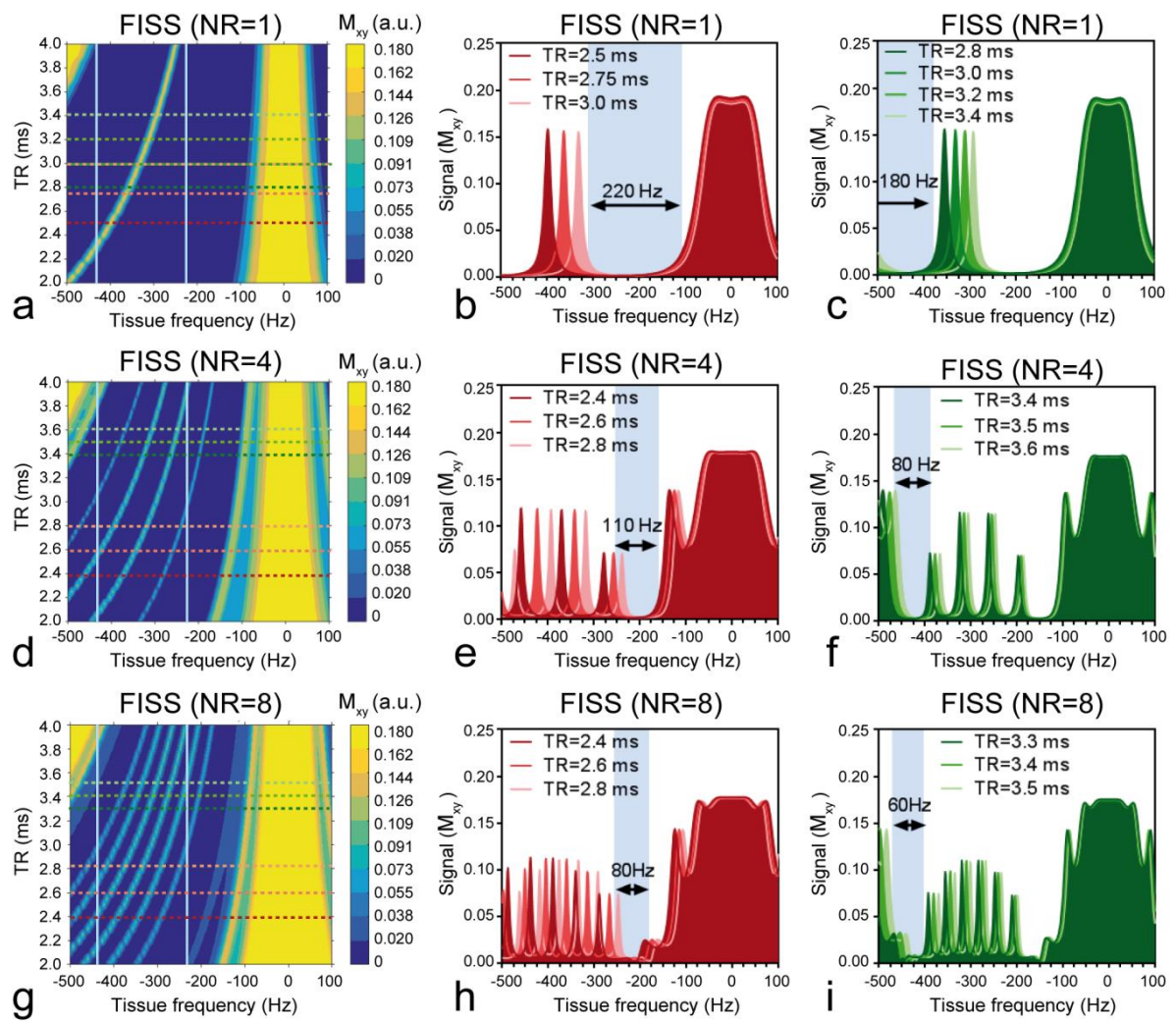

*Figure 3*

A fat profile consisting of a normalized gaussian distribution with an FWHM of 50Hz around -440Hz and -220Hz was used to simulate the effect of TR and NR on the transverse magnetization (a,d). These plots were normalized to the maximum on-resonance (i.e. water) transverse magnetization. Line plots demonstrate clearly the effect of NR on fat suppression as function of TR compared with bSSFP (b,e) and this signal behavior was corroborated by the experimental fat SNR observed in a baby oil phantom (c,f). The darkest blue region (in a, d) correspond to the bottom 10% of Mxy and corresponds to the region where fat is considered suppressed.

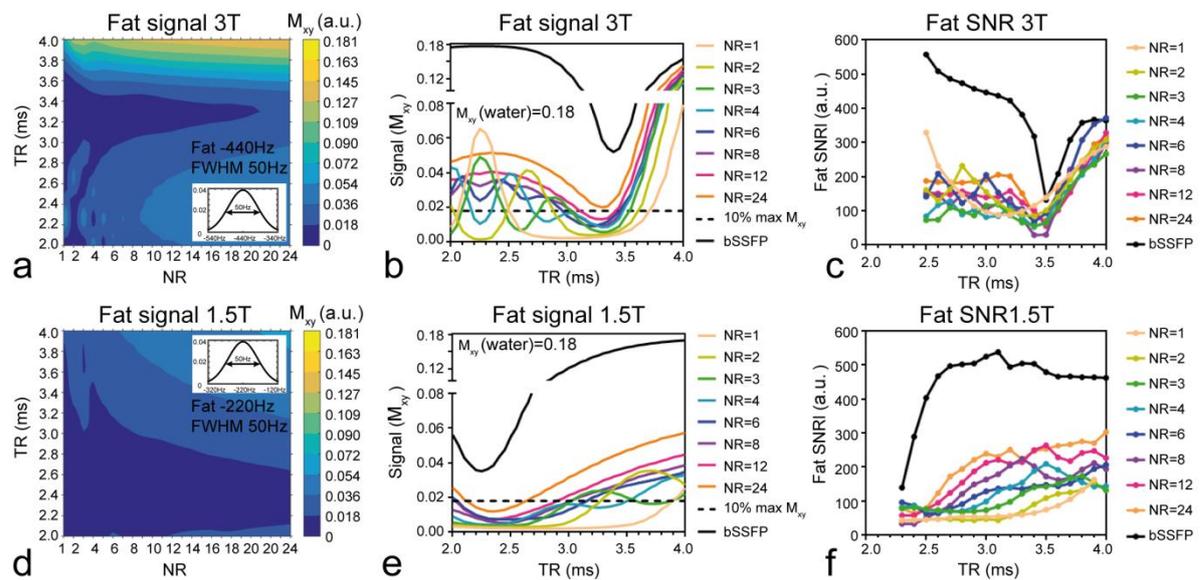

*Figure 4*

Comparison of 5D whole-heart images multiplanar reformatted in sagittal planes and acquired with free-breathing non ECG-triggered FISS (NR=8) and bSSFP at 3T. Images are displayed in systole and mid-diastole, at both end-expiratory and end-inspiratory phases. Yellow lines serve to indicate the respiratory motion. Note the decrease in streaking artifacts in FISS compared with bSSFP indicated by the red arrows. The scan duration was 5:49min for FISS (NR=8, TR=3.4ms) and 4:39min for bSSFP for an isotropic resolution of 1.38mm$^3$. Motion can be better appreciated in an animation (see Supporting Figure 4).

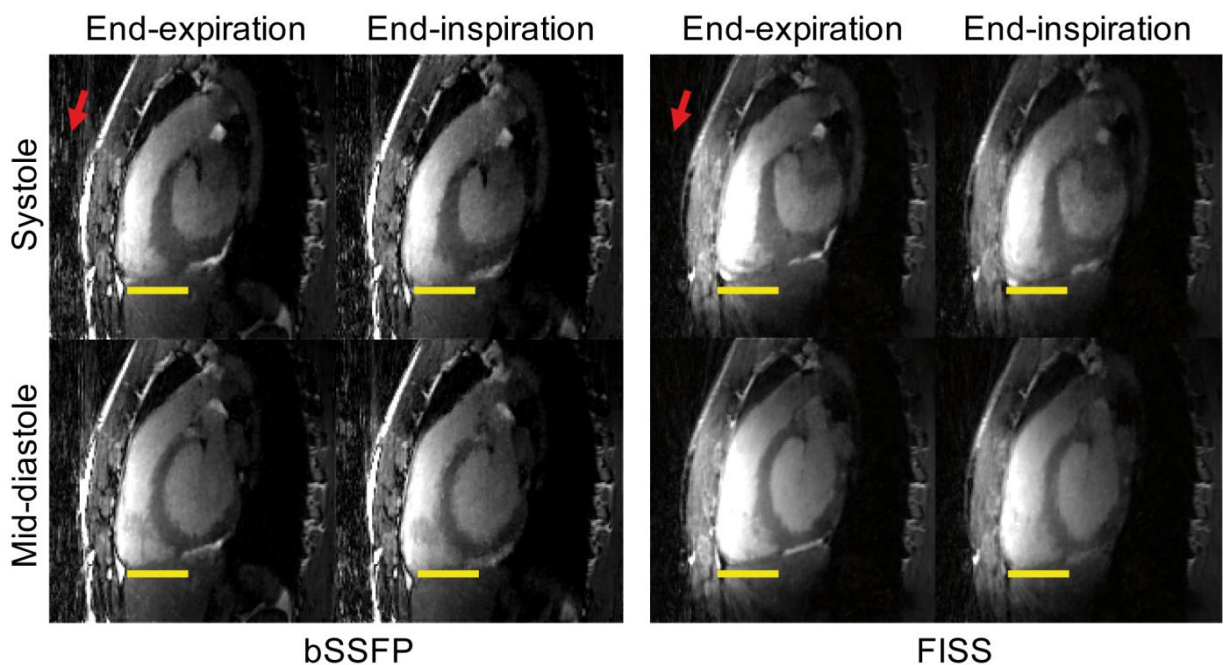

*Figure 5*

Comparison of 5D whole-heart images in coronal planes acquired with free-breathing non ECG-triggered FISS and bSSFP at 1.5T. In this subject data was reconstructed using 16 cardiac phases and 4 respiratory phases. The displayed images correspond to the end-expiratory phase and only 8 out of 16 cardiac phases are shown. Both respiratory and cardiac motion can be better appreciated in an animation (Supporting Figure 5). Scan time duration was 4:46 min for FISS (NR=8, TR=2.75ms) and 3:45min for bSSFP for an isotropic resolution of 1.38mm$^3$. Red arrows indicate cardiac regions containing fat that is suppressed with FISS but not with bSSFP.

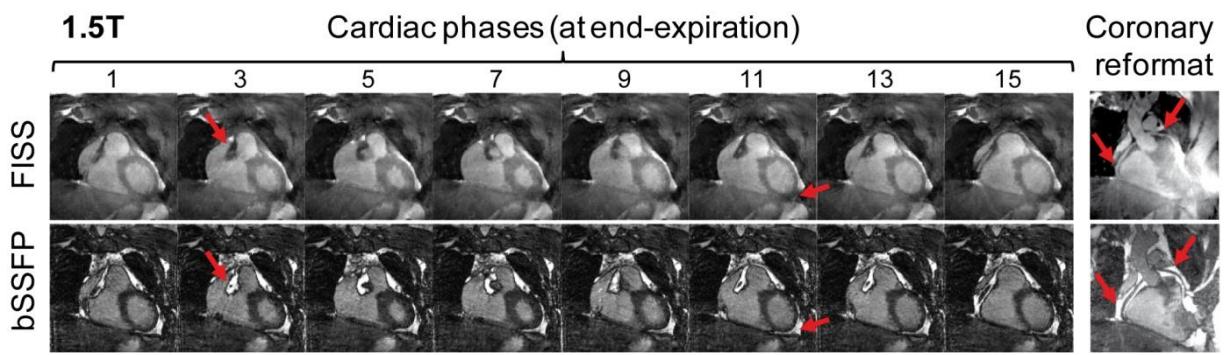

*Figure 6*

Comparison of 5D whole-heart images in coronal planes acquired with free-breathing non ECG-triggered FISS and bSSFP at 3T. In this subject data was reconstructed using 26 cardiac phases and 4 respiratory phases. Only 7 out of 26 cardiac phases are shown, all at end expiration. Both respiratory and cardiac motion can be better appreciated in an animation (Supporting Figure 5). Scan time duration was 6:20min for FISS (NR=4, TR=3.0ms) and 4:08min for bSSFP for an isotropic resolution of $1.38mm^3$. Red arrows indicate cardiac regions containing fat that is suppressed with FISS but not with bSSFP.

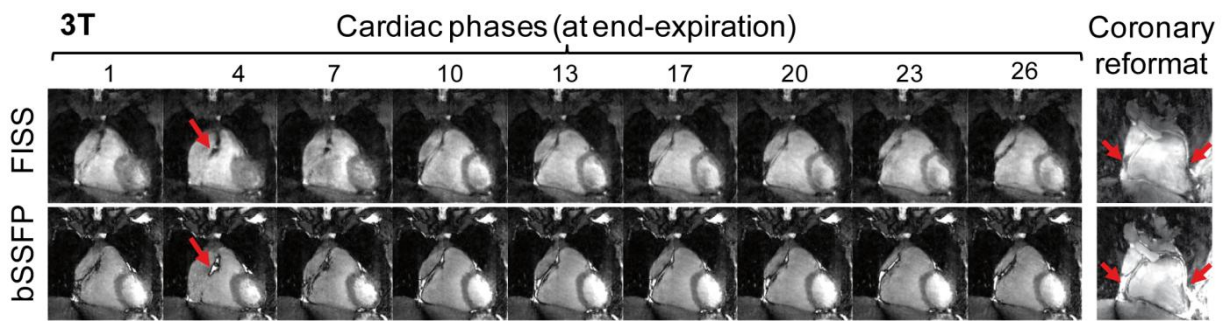

*Figure 7*

Comparison of coronary reformats obtained with FISS and bSSFP with an isotropic resolution of 1.1mm$^3$ at 1.5 and 3T (a). The water-fat cancellation artifacts at the coronary vessel borders can be clearly observed in the bSSFP scans. Therefore a quantitative comparison of vessel sharpness across techniques was not made as the dark delineation of the coronary arteries in the bSSFP scans would bias the results. The scan time duration at 1.5T was 7:53min for FISS (NR=8, TR=3.0ms) and 6:19min for bSSFP. At 3T, the scan time duration was 8:45min for FISS (NR=8, TR=3.4ms) and 7:00min for bSSFP. The SNR of fat was determined in non-motion-resolved images and was significantly decreased in FISS compared with bSSFP.

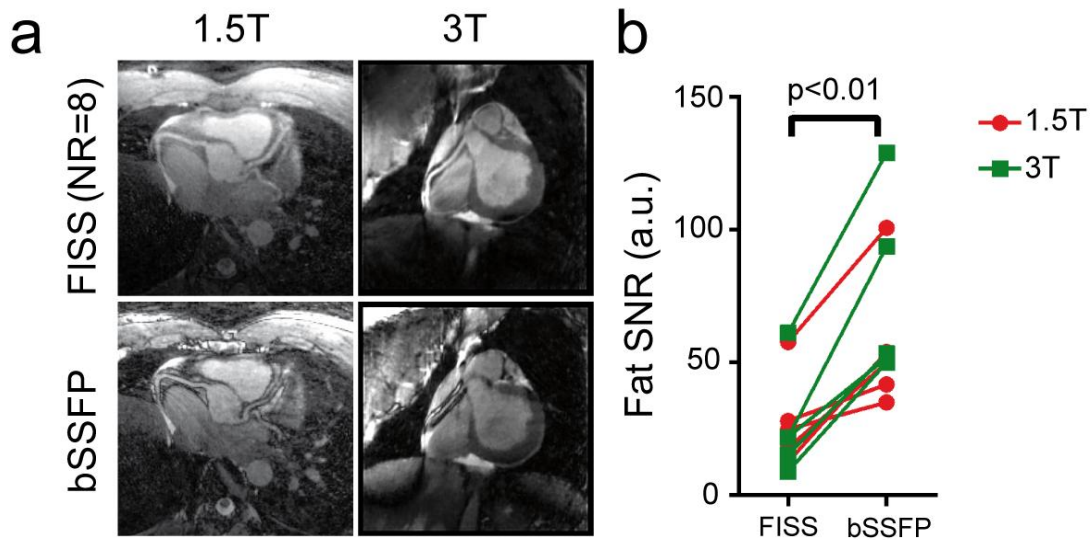

*Figure 7*

## SUPPORTING FIGURE CAPTIONS

*Supporting Figure S1*

Bloch simulations of bSSFP (a) and FISS (NR=1) (b) were performed to compare the transverse magnetizations as function of tissue frequency and RF excitation angle. Both with bSSFP and FISS, the top 10% of $M_{xy}$ can be obtained within a similar RF excitation angle range. FISS provides the same maximum on-resonant (~0Hz) signal intensity compared with bSSFP. However, a broad fat suppression bandwidth with NR=1 can be observed around -220 Hz and -440 Hz, while bSSFP displays a narrow band, occurring as expected at frequencies of 1/TR.

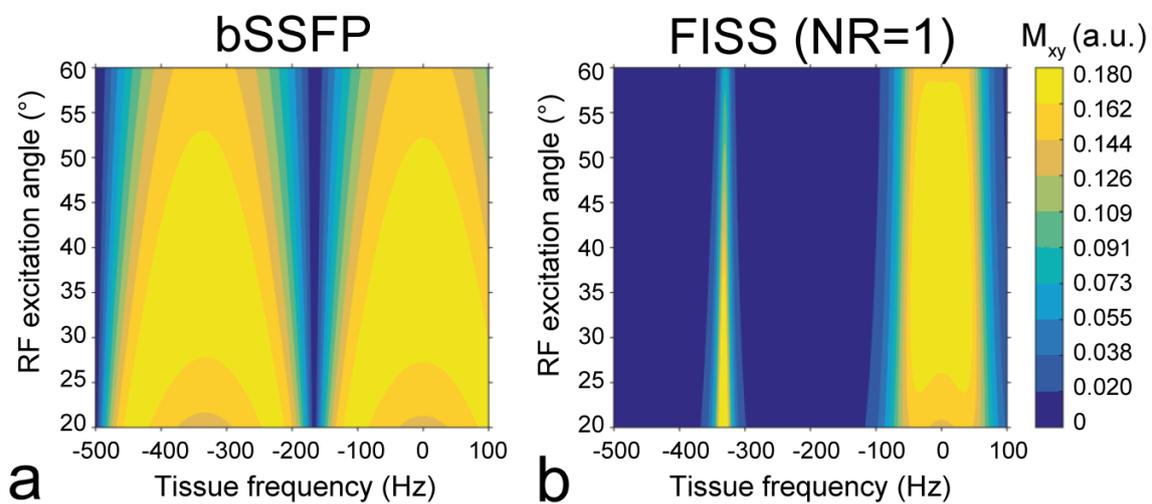

## *Supporting Figure S2*

Bloch simulations of FISS illustrate the effect on the transverse magnetization as function of TR and tissue frequency. For NR=1 to NR=24. An increase in the number of readouts per FISS module (NR) mainly affects the width of the fat suppression bands, while a change in TR mainly affects the location of the bands. The white arrows indicate regions of fat suppression for either -440Hz or -220Hz which corresponds to the frequency of fat at 3T or 1.5T respectively.

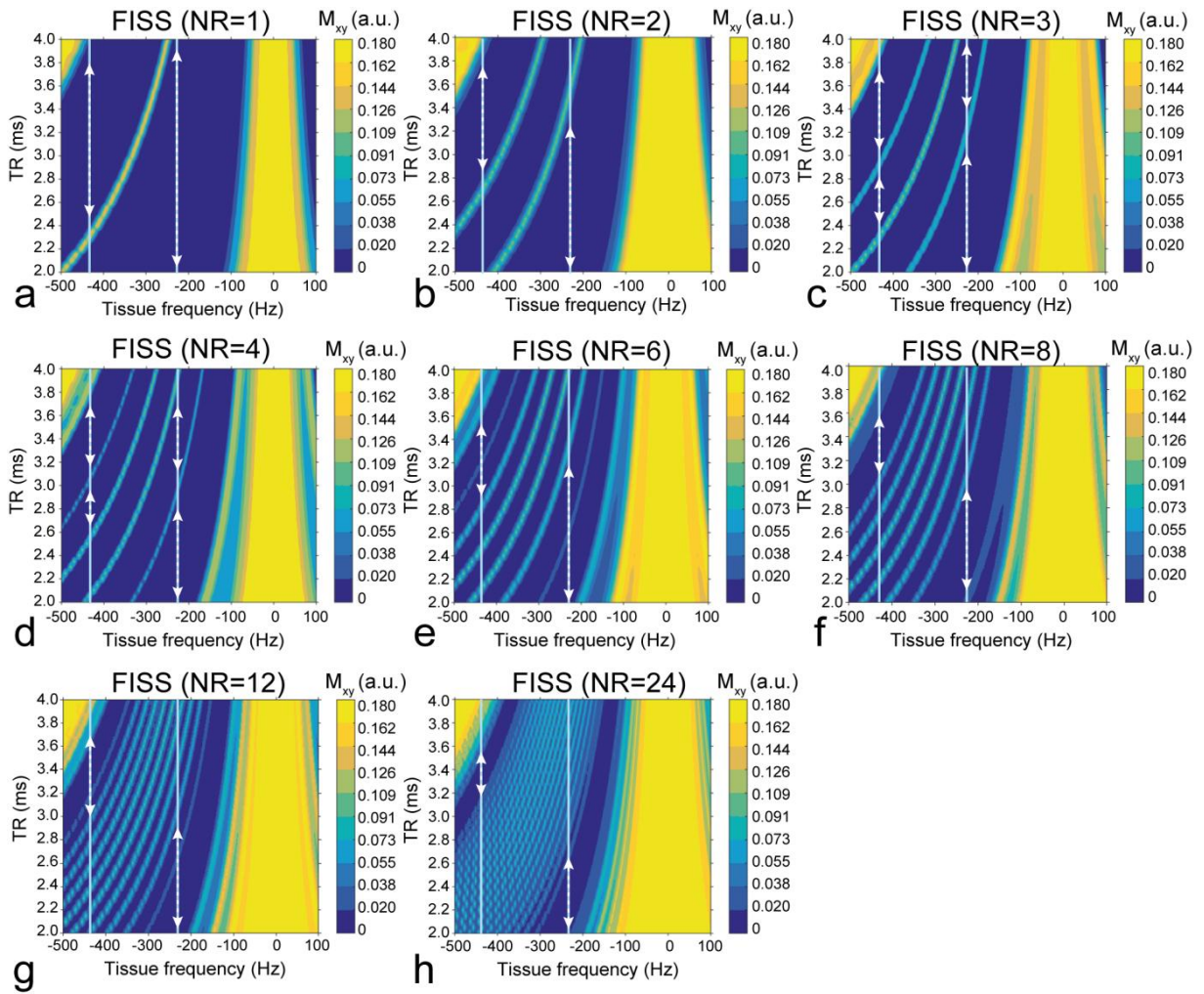

### *Supporting Figure S3 (Animation)*

The image quality of a 3D FISS acquisition in a phantom is similar with and without the use of gradients spoilers (a,b). However the SI projections are not consistent without using gradient spoilers (c). Applying gradient spoilers stabilizes the SI projections (d) which is important for motion signal extraction.

### *Supporting Figure S4 (Animation)*

Animation showing cardiac and respiratory motion from data acquired with free-running FISS and bSSFP. Data corresponds to the same volunteer images shown in Figure 4 of the manuscript.

### *Supporting Figure S5 (Animation)*

Animation showing cardiac and respiratory motion from data acquired with free-running FISS and bSSFP. Data corresponds to the same volunteer images shown in Figure 5 and 6 of the manuscript.